# MIRST-DM: Multi-Instance RST with Drop-Max Layer for Robust Classification of Breast Cancer

Shoukun Sun, Min Xian, Aleksandar Vakanski, Hossny Ghanem

Department of Computer Science, University of Idaho, Idaho Falls, USA

**Abstract.** Robust self-training (RST) can augment the adversarial robustness of image classification models without significantly sacrificing models' generalizability. However, RST and other state-of-the-art defense approaches failed to preserve the generalizability and reproduce their good adversarial robustness on small medical image sets. In this work, we propose the Multi-instance RST with drop-max layer, namely MIRST-DM, which involves a sequence of iteratively generated adversarial instances during training to learn smoother decision boundaries on small datasets. The proposed drop-max layer eliminates unstable features and helps learn representations that are robust to image perturbations. The proposed approach was validated using a small breast ultrasound dataset with 1,190 images. The results demonstrate that the proposed approach achieves state-of-the-art adversarial robustness against three prevalent attacks.

## 1   Introduction

Deep neural networks (DNNs) have achieved unprecedented performance in medical image analysis; however, existing approaches lack robustness to both natural and adversarial image perturbations [1], which makes it difficult to deploy them in an open-world environment. In this work, we focus on improving the adversarial robustness of deep learning-based models for medical image classification. Varies defense approaches were developed to overcome DNNs' vulnerability to adversarial samples by incorporating adversarial samples into training to improve the smoothness of a model's decision boundary [1–6]. Existing approaches were proposed on natural image sets, and no study has been conducted to improve the adversarial robustness of medical systems. Most natural image sets are large, e.g., the CIFA-10 dataset has 6k images, and the ImageNet has more than 14 million images, while typical medical image datasets [7, 8] are usually much smaller than natural image datasets. The small medical datasets make it difficult for adversarial training to converge to a smooth decision boundary; hence established approaches cannot reproduce their good adversarial robustness on medical image sets. Furthermore, previous research showed that a tradeoff exists between the adversarial robustness and generalization of a DNN [4]; and some defense approaches [1] have a dramatic decrease in their generalizability.

To overcome the challenges in the existing defense approach, we propose the multi-instance robust self-training with the drop-max layer to augment DNNs' adversarial robustness on small medical image sets. We are inspired by the finding that training with a series of adversarial samples could lead to smoother decision boundaries and achieve higher adversarial robustness. The proposed approach utilizes multiple adversarial instances from a sequence of images generated iteratively from a multi-step attack; and it increases the number of training samples and encodes the smooth transitions among adversarial samples. Furthermore, we propose the drop-max layer to replace the first pooling layer to remove unstable features at an early stage. We validate different approaches using a small breast ultrasound dataset [7, 8]. The primary contributions of this study are summarized below.

- The proposed approach uses a sequence of adversarial samples during training and significantly improves the adversarial robustness of the conventional RST approach using a small image set.
- The proposed drop-max layer removes non-robust features and greatly improves the adversarial robustness of RST approaches against the Carlini and Wagner attack and Projected Gradient Descent attacks.
- The pretrained backbone network using contrast learning improves both the generalization and adversarial robustness of a DNN model.

## 2 Related Works

### 2.1 Adversarial Attacks and Defenses

Adversarial attacks can be categorized into white-box and black-box attacks. A white-box attack generates adversarial examples directly using parameters from the target neural network; on the contrary, a black-box attack is applied on a surrogate model instead of accessing the actual target model. Attacks can be targeted and untargeted. Targeted attacks fool models to produce intended classification output, while untargeted attacks generate adversarial examples that can be misclassified to any class which differs from the true label. In this study, we focus on defending targeted white-box attacks.

Fast Gradient Sign Method (FGSM) [1] computes the gradients of a loss function with respect to the target image, then uses the gradients as perturbations and adds them to each pixel. FGSM is also called the one-step first-order attack because it only calculates the first-order gradient once to generate adversarial examples. Projected Gradient Descent (PGD) [9] iteratively updates perturbations based on the gradients with respect to the adversarial image by $n$ steps. In each loop, PGD works like FGSM but with a smaller step size while updating perturbations. After each loop, it cuts perturbation values that are larger than the allowed maximum perturbation $\epsilon$. Carlini and Wagner (CW) [10] has three implementations: $L_0$, $L_1$, and $L_\infty$, and we use the $L_\infty$ version in this study. Instead of using cross-entropy loss function, CW applies a new loss function to find stronger adversarial examples and produces smaller perturbations. The framework of CW with $L_\infty$ setting is similar to the PGD attack excepting the loss function.

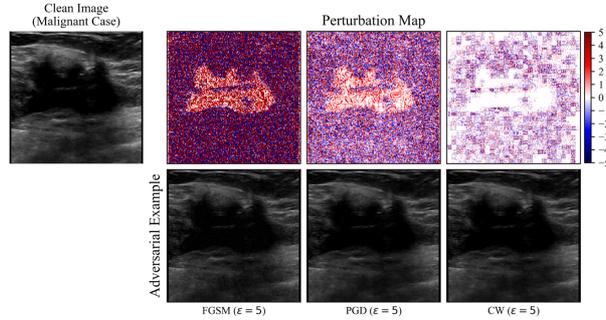

**Fig. 1.** Adversarial samples. The first row presents perturbations generated by three attacks. The second row shows the corresponding adversarial samples. The adversarial examples are misclassified to benign cases.

The two most popular defense approaches are regularization and training with adversarial samples. Adversarial Training (AT) [9] trains DNNs with adversarial examples generated with the model parameters on the current epoch. Classifiers trained with AT could increase robustness, but it costs a significant decrease of a model's generalization ability. Gradient Regularization (GR) [2] introduces the $L_p$ norm of gradients with respect to the input into the loss function. The model trained with the additional gradient term could produce smoother classification boundaries. GR conducts double backpropagation and needs more computations and memory. Locally-Linear Regularization (LLR) [3] regularizes the classifier to be locally linear using the first-order Taylor expansion of a loss function. Tradeoff-inspired Adversarial Defense via Surrogate-loss minimization (TRADES) [4] is motivated to reduce the tradeoff between standard and adversarial accuracies using a classification-calibrated surrogate loss as the regularization term. The TRADES solves the problem of AT by adding a soft regularization term to minimize the difference of outputs between the clean image and the crafted image given. Robust Self-Training (RST) [5][6] was first proposed to utilize unlabeled images with the semi-supervised learning framework to improve the adversarial robustness. For making fair comparison with other defense approaches, we only consider the supervised RST which uses only labeled images and a loss function with the standard cross-entropy loss and the adversarial loss.

### 2.2 Breast Ultrasound Image Classification

Breast ultrasound (BUS) was recommended as a primary imaging tool or supplemental screening with mammography to improve the sensitivity of cancer detection in women with dense breasts [11–13]. Recent research [14–17] demonstrated that BUS images with advanced DNNs could achieve high sensitivity and specificity for breast cancer detection. Shia et cl. [14] extracted meaning features from BUS images using a pre-trained ResNet-101 followed by a linear SVM classifier. The sensitively used images from 543 patients could reach 93.94%. Xie et al. [15] proposed the dual-sampling network by combining a conventional convolutional network and a residual network. The approach achieved a test accuracy of 91.67% on a dataset with 1,272 BUS images. Zhang et al. [17] proposed a multitask DNN to segment and classify breast tumor in BUS images simultaneously. The approach was validated using 647 images and achieved a test accuracy of 95.56%. However, an approach's excellent performance on BUS images from one image set often degrades significantly on images from a different set, or even on images from the same set with minor perturbations. This indicates the lack of robustness of existing approaches which undermines the confidence of clinicians in adopting these computer-aided diagnosis systems.

## 3    Proposed Method

The previous work [6] demonstrated that the adversarial robustness of deep learning models could be improved with the increasing number of training samples, while it is difficult to achieve high generalization and adversarial robustness using small training sets. The effectiveness of existing approaches was usually validated using large datasets with 10k to 100k or even more samples. However, many typical medical image sets [7, 8] only have a few hundred to thousand images, and applying previous approaches, e.g., TRADES and RST, achieves poor robustness. Inspired by RST, we propose a novel self-training approach that applies the newly proposed drop-max layer and involves multiple adversarial samples during training to smooth the decision boundaries.

### 3.1    Multi-Instance RST

Let $\{(x_i, y_i)\}_{i=1}^{n}$ be a labeled dataset, and $f_\theta$ be a parametric model to be learned during the training to map $x_i$ to $y_i$. The general RST approach for supervised learning aims to minimize a combination of a standard loss ($L_{std}$) and an adversarial loss ($L_{adv}$),

$$L_\theta = \underbrace{\frac{1}{n}\sum_{i=1}^{n} l(f_\theta(x_i), y_i)}_{L_{std}} + \underbrace{\frac{1}{n}\sum_{i=1}^{n} \max_{\bar{x} \in D(x_i)} l(f_\theta(\bar{x}), y_i)}_{L_{adv}} \qquad (1)$$

where $l(\cdot)$ denotes a standard loss function used in model training, e.g., the cross-entropy function; and $D(x_i)$ is a set of adversarial samples of $x_i$.

In Eq. (1), $L_{adv}$ only counts the most aggressive adversarial instances from $D(x_i)$ for each clean image. When there is a large difference between $x$ and $\bar{x}$, it is difficult for a model to converge to a smooth classification boundary that encloses both $x$ and the most aggressive $\bar{x}$ using a small number of training images. Our motivation is to enable the visibility of the smooth transition from a clean image to the most aggressive adversarial instance during the training. We propose the Multi-instance RST (MIRST) to overcome existing challenges by involving a sequence of adversarial instances to speed the model convergence and enhance the adversarial robustness. The loss function of the proposed MIRST is defined by

$$L_\theta^{MI} = \frac{1}{n}\sum_{i=1}^{n} l(f_\theta(x_i), y_i) + \alpha \cdot \frac{1}{n}\sum_{i=1}^{n}\sum_{j=1}^{m} \beta_j \cdot l(f_\theta(\bar{x}_j), y_i) \qquad (2)$$

where $\alpha$ balances the standard loss and the adversarial loss, $\beta_j$ denotes the weight for the loss of the $j$th adversarial instance $\bar{x}_j$, and $m$ is the number of adversarial instances.

In this work, we applied a T-step PGD attack to generate adversarial instances,

$$x_{t+1}^{adv} = \Pi_{x\pm\epsilon}\left(x_t^{adv} + \gamma \cdot sign\left(\nabla_x l(f_\theta(x_t^{adv}), y)\right)\right), t = 1, 2, \cdots, T \quad (3)$$

where $x_t^{adv}$ is the $t$th adversarial instance and $x_1^{adv}$ is the original clean image; the $sign(\cdot)$ function returns the sign of a value; $\Pi_{x\pm\epsilon}(\cdot)$ operator projects values to the range $[x - \epsilon, x + \epsilon]$; and $\gamma$ is the step size. In experiments, T is set to 10, and $\bar{x}_j$ is $x_{2j+1}^{adv}$.

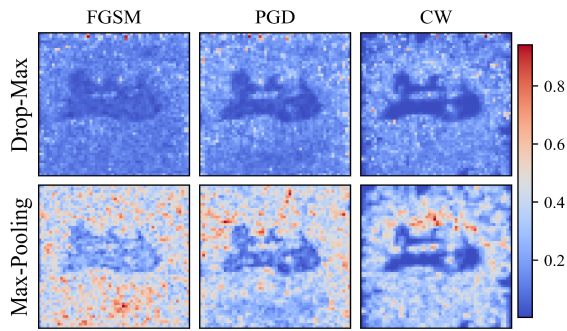

**Fig. 2.** Different residual maps calculated using the max-pooling layer and drop-max layer. Each residual map shows the difference between a clean image's feature map and its adversarial image's feature map. Note that all channels are added to generate residual maps.

### 3.2 Drop-Max Layer

The max-pooling is a popular operation in convolutional neural networks (CNNs) to downsample feature maps and extract the most salient features. However, those maximum values are sensitive to image perturbations. To overcome this challenge, we propose the drop-max layer to remove unstable features at the early stage of CNNs.

The drop-max layer removes the maximum value and selects the second largest value into the next layer. Most white-box attacks utilize the gradient $\nabla_x \mathcal{L}(f_\theta(x), y)$ to find perturbations that fool the model to produce the wrong output. Because of the max-pooling layers, perturbations were transmitted to the final decision layer through the unstable maximum values. We replace the first max-pooling layer with the proposed drop-max layer to depress the chance of letting perturbed values get into the next layers.

### 3.3 SimCLR Pretraining

The SimCLR [18] is a semi-supervised learning framework that was proposed to learn high-quality image representations using a small labeled image set. SimCLR samples a minibatch with the size of $N$ and pair each image with an augmented image. By treating each pair as a positive pair, the rest $2(N-1)$ images are treated as negative. The normalized temperature-scaled cross-entropy (NT-Xent) is used to calculate loss across all positive pairs,

$$l_{simclr}(z_i, z_j) = -\log \frac{\exp\left(\text{sim}(z_i, z_j)/\tau\right)}{\sum_{k=1}^{2N} \mathbb{1}_{[k \neq i]} \exp(\text{sim}(z_i, z_k)/\tau)}, \qquad (4)$$

where $z_i$ and $z_j$ are the outputs two image samples in a positive pair, $\text{sim}(\cdot)$ is the cosine similarity function, $\mathbb{1}_{[k \neq i]}$ is an indicator function that returns 1 iff $k \neq i$, and $\tau$ is a constant. Minimizing $l_{simclr}$ leads to similar outputs of images in positive pairs and increased differences in outputs between positive and negative images.

## 4 Experimental Results

### 4.1 Experiment Setup

**Datasets and metrics**. We apply two B-mode BUS image sets, BUSI [7] and BUSIS [8], to evaluate the proposed method. The combined dataset has 1,190 images, of which 726 are benign and 464 are malignant. All images are resized to 224×224 pixels. To avoid the distortion of tumor shape, we apply zero-padding to generate square images before image resizing. The 5-fold cross-validation with F1-score and sensitivity are used to quantitatively evaluate the adversarial robustness of different defense strategies.

**Network and training**. ResNet-50 [19] is the most popular network used to validate the effectiveness of different defense strategies [9, 20, 21], and we use it as the baseline image classifier in all experiments. All approaches are trained for 100 epochs using the Adam optimizer with a learning rate of 0.0001 and a batch size of 8.

**Adversarial attack settings**. We employ three adversarial attack approaches, FGSM [1], PGD [9], and CW [10]. All these attacks are bounded by $\epsilon = 0.0196$ (5/255) with respect to the $L_\infty$ norm. The number of the maximum iterations of PGD and CW is set to 20 and 10, respectively, and the step sizes are both 0.001.

**Defense methods and parameters**. We implement the AT [9], GR [2], LLR [3], TRADES [4], and RST [5, 6]. The weight of adversarial loss in GR, TRADES, and RST is set to 1 by experiments. In LLR, the weight of the locally linearity measure and gradient magnitude term is set to 3 and 4, respectively. MIRST-DM uses weight 1 for the adversarial loss for a fair comparison with RST. The weights for the four adversarial examples in MIRST-DM follow a descending series, i.e., $\beta = [0.34, 0.28, 0.22, 0.16]$, which sums to be 1.

### 4.2 The Effectiveness of Multiple-Instance RST

The primary difference between RST and the proposed MIRST is the number of adversarial samples involved in the training. In this section, we compare their performance against three attack methods. ResNet50 is used as the image classifier for the baseline, RST, and MIRST and pre-trained using ImageNet. The baseline model uses the CE loss. The adversarial robustness is evaluated using the F1 score and sensitivity on adversarial images from three attacking approaches. Note that all metrics listed in Tables 1-3 are the average values of the results from 5-fold cross-validation.

**Table 1.** Test performance of the baseline model, RST, and MIRST.

| | Attack | Baseline | RST | MIRST |
|---|---|---|---|---|
| F1 score | No attack | **0.798** | 0.757 | 0.748 |
| | FGSM | 0.014 | 0.519 | **0.542** |
| | PGD | 0.000 | 0.496 | **0.516** |
| | CW | 0.000 | 0.482 | **0.498** |
| Sensitivity | No attack | **0.786** | 0.711 | 0.713 |
| | FGSM | 0.017 | 0.495 | **0.528** |
| | PGD | 0.000 | 0.478 | **0.513** |
| | CW | 0.000 | 0.463 | **0.492** |

In Table 1, the values in the 'No attack' row are calculated using clean test images, while values in other rows are from adversarial images. The F1 score and sensitivity of the baseline model drop significantly under the three attack approaches, which demonstrates the poor adversarial robustness of the baseline model. For example, the PGD and CW attacks reduce the baseline model's F1 score to 0. Both RST and MIRST achieve comparable generalization and greatly improve the adversarial robustness of the baseline model against the FGSM, PGD, and CW attacks. MIRST achieves higher F1 scores than those of RST against all three attacks. MIRST improves RST's sensitivity values against the FGSM, PGD, and CW attacks by 6.74%, 7.17%, and 6.22%.

**Table 2.** Test performance of the models using SimCLR pretrained model.

| | Attack | Base-line | GR | AT | LLR | TRADES | RST | MI-RST |
|---|---|---|---|---|---|---|---|---|
| F1 score | No attack | 0.831 | 0.817 | 0.489 | - | 0.822 | 0.802 | **0.830** |
| | FGSM | 0.231 | 0.198 | 0.253 | - | 0.576 | 0.620 | **0.640** |
| | PGD | 0.000 | 0.000 | 0.096 | - | 0.366 | 0.425 | **0.544** |
| | CW | 0.000 | 0.000 | 0.479 | - | 0.458 | 0.467 | **0.512** |
| Sensitivity | No attack | 0.827 | 0.815 | 0.344 | - | 0.780 | 0.762 | **0.805** |
| | FGSM | 0.284 | 0.245 | 0.267 | - | 0.522 | 0.604 | **0.622** |
| | PGD | 0.000 | 0.000 | 0.083 | - | 0.379 | 0.420 | **0.543** |
| | CW | 0.000 | 0.000 | 0.336 | - | 0.467 | 0.463 | **0.504** |

'-' in the LLR column denotes unavailable values because the training did not converge.

### 4.3 The Effectiveness of The SimCLR Pretrained Model

We compare the generalization ability and adversarial robustness of the baseline, GR, AT, LLR, TREADS, RST, and MIRST pretrained using the SimCLR loss on ImageNet. All results in Table 1 are from models pretrained using the CE loss on ImageNet. As shown in Table 2, SimCLR improves the generalizability (F1 scores) of the baseline,

RST, and MI-RST by 4.14%, 5.87%, and 10.93%. In addition, SimCLR can significantly improve the adversarial robustness of the proposed MIRST, e.g., MIRST's F1 scores against FGSM, PGD, and CW attacks increased by 18.06%, 5.46%, and 2.84%.

**Table 3.** Results of RST and MI-RST with (w/) and without (w/o) the drop-max layer.

| | | RST | | MIRST | |
|---|---|---|---|---|---|
| | Attack | w/o DM | w/ DM | w/o DM | w/ DM |
| F1 score | No attack | 0.802 | **0.825** | **0.830** | 0.823 |
| | FGSM | 0.620 | **0.641** | 0.640 | **0.655** |
| | PGD | 0.425 | **0.580** | 0.544 | **0.586** |
| | CW | 0.467 | **0.740** | 0.512 | **0.734** |
| Sensitivity | No attack | 0.762 | **0.806** | **0.805** | 0.799 |
| | FGSM | 0.604 | **0.628** | 0.622 | **0.650** |
| | PGD | 0.420 | **0.571** | 0.543 | **0.587** |
| | CW | 0.463 | **0.743** | 0.504 | **0.720** |

### 4.4 The Effectiveness of The Drop-Max Layer

We compare the results of the RST and MIRST with and without the proposed drop-max (DM) layer. The DM layer is applied to replace the first max-pooling layer of ResNet50. As shown in Table 3, the F1 scores of RST with the DM layer against the FGSM, PGD, and CW are improved by 3.38%, 36.66%, and 58.51%, respectively. The DM layer also increases the F1 score of RST on clean images by 2.89%. The DM layer does not improve MIRST's generalization on clean images. But it improves the adversarial robustness of MIRST against all three attacks. E.g., the F1 score of MIRST-DM against the CW attack is 43.42% higher than that of MIRST without the DM layer; and the F1 score of MIRST-DM against the PGD attack increased by 7.65%.

## 5 Conclusion

In this study, we propose the MI-RST approach to improve the adversarial robustness of DNNs and validate it using a breast ultrasound (BUS) image set. The proposed MI-RST achieves higher adversarial robustness than RST using a small BUS image set, and its generalizability is comparable to RST. The proposed drop-max layer removes unstable features at the early stage of model training and significantly improves the adversarial robustness of both RST and MI-RST against FGSM, PGD, and CW attacks. The pretrained model using SimCLR leads to improved generalizability and adversarial robustness of MI-RST. This work provides a comprehensive view of the performance of state-of-the-art defense approaches on a small medical image set. It can help develop more robust and trustworthy computer-aided diagnosis systems in an open-world environment.